\documentstyle[prl,aps,epsf,multicol]{revtex}
\begin{document}
\title{Temperature-dependent logarithmic corrections in the spin-$1/2$  
Heisenberg chain.}
\author{Victor Barzykin}
\address{National High Magnetic Field Laboratory, 
Florida State University,\\
1800 E. Paul Dirac Dr., Tallahassee, Florida 32310}
\maketitle
\begin{abstract}
We obtain the logarithmic corrections to the
dynamic response function and NMR $T_1$ and $T_{2G}$ rates
in the spin-$1/2$ antiferromagnetic Heisenberg chain using
perturbative renormalization group in the leading irrelevant operator.
The result is compared with NMR experiments in Sr$_2$CuO$_3$.
\end{abstract}
\begin{multicols}{2}
\narrowtext
Quantum spin chains have attracted considerable interest for a long
time, due to both unconventional physics of the $1D$ materials and
sophisticated theoretical methods used in the analysis of the problem.
In particular, it is well known\cite{Affleck:review} that the 
antiferromagnetic spin-$1/2$ $XXZ$ chain is critical for $J_x>J_z$
(the two exchange constants),
and the spin correlators at $T=0$ decay as a power law with distance.
The critical theory can be mapped on a theory of free bosons.
For $J_x<J_z$ the system acquires a spin gap. 
At the quantum critical point, 
the Heisenberg $XXX$ chain, the spin 
correlation function has logarithmic corrections to the
free theory coming from the leading irrelevant operator, which
becomes marginal.
Thus, the asymptotic power law behavior of the correlator at long 
distances is modified\cite{Affleck,Singh}. The staggered component
of the spin-spin correlator has the following form:
\begin{equation}
\langle S^z(r) S^z(0) \rangle = 
(-1)^r {\sqrt{\ln(r/r_0)} \over (2 \pi)^{3/2} r},
\label{infcor}
\end{equation}
where the non-universal coefficient $D=1/(2 \pi)^{3/2}$ has been determined 
from the Bethe Anzats\cite{Affleck1,Lukyanov}.
 
Although this logarithmic behavior  has been known for some 
time\cite{Affleck,Singh},
numerical\cite{Kubo,Liang,Sandvik,Hallberg,Koma} and experimental
\cite{neutrons,takigawa} tests of it have appeared 
only recently. To fit the numerical data, various anzatses
were used\cite{Kubo,Hallberg,Koma,starykh}. 
A consistent field-theoretical derivation of the logarithmic factor for 
the equal-time correlator in a finite-size chain at $T=0$ has 
been recently done in Ref.\cite{Affleck1,BA}. In what
follows we use a similar approach to obtain staggered dynamic spin correlator,
up to two-loop order in the perturbative renormalization group scheme. 
The imaginary part of the staggered dynamic spin susceptibility  is 
observed in the inelastic neutron scattering experiments, and
determines NMR $T_1$. NMR $T_{2G}$ is found from the real part 
of the spin susceptibility.

Let us now show how the logarithms appear in the time-dependent staggered spin
correlation function. For this purpose it is more convenient to work
in real space and imaginary time, so that one can easily apply bosonization
and conformal invariance. One should then Fourier transform and analytically
continue the result to real frequences.
At low temperatures and large distances we can use the continuum
approximation. The theory can be written in terms of free bosons 
defined on a circle.  In case of $SU(2)$ - symmetric Heisenberg
model it is more convenient to use the  non-Abelian bosonization\cite{Affleck},
which respects the symmetry. The
action for the $SU(2)$-symmetric matrix field ${\bf g}^{\alpha}_{\beta}$ 
includes the Wess-Zumino term with coefficient $k=1$.
The Hamiltonian density takes the following form:
\begin{equation}
H = H_0 - 8 \pi^2/\sqrt{3} \lambda {\bf J}_L \cdot {\bf J}_R,
\label{hamilt}
\end{equation}
where $H_0$ is the Hamiltonian density for a free boson, ${\bf J}_{L,R}$ 
are left and right $SU(2)$ currents:
\begin{equation}
{\bf J}_L \equiv { -i \over 4 \sqrt{\pi}} tr[{\bf g}^{\dagger} 
\partial_-{\bf g}\bbox{\sigma}], \ \ \
{\bf J}_R \equiv { i \over 4 \sqrt{\pi}} tr[\partial_+{\bf g} 
{\bf g}^{\dagger}\bbox{\sigma}]
\end{equation}

The spin operators can be written in non-abelian bosonization notation as:
\begin{equation}
{\bf S}_j  = ({\bf J}_L + {\bf J}_R) + \hbox{const}\ i 
(-1)^j tr[{\bf g} \bbox{\sigma}],
\end{equation}
so that the correlation function has uniform and staggered terms,
\begin{equation}
\chi(r,\tau)=\langle S_0^z S_r^z \rangle \rightarrow \chi_u(r,\tau) + (-
1)^r \chi_s(r,\tau),
\end{equation}
where $\tau$ is imaginary time.
Both terms vary slowly on the scale of the lattice spacing, and
correspond to different Green's functions in the 
continuum theory. The staggered susceptibility, which is enhansed near 
$q=\pi$ wave vector, is observed by inelastic neutron scattering and NMR,
\begin{equation}
\label{stc}
\chi_s(r,\tau) \propto 
\langle tr(\bbox{\sigma}^z {\bf g})(r,\tau) tr(\bbox{\sigma}^z 
{\bf g})(0)\rangle.
\end{equation}

It is not difficult to determine the contribution of the free boson with
radius $R=1/\sqrt{2 \pi}$\cite{Tsvelik} - the conformally invariant WZW model on a circle
of length $\beta = 1/T$ in the imaginary time direction. Indeed, 
${\bf g}$ has scaling dimension 1/2, so for an infinite system one writes:
\begin{equation}
<tr [{\bf g}(z,\bar{z})\bbox{\sigma}^z]tr [{\bf g}(0)\bbox{\sigma}^z]>={1\over
\sqrt{z\bar{z}}}.\label{gcorr}\end{equation},
where $z=\tau + ix$, $\bar{z}=\tau - ix$.
Here we have chosen a convenient normalization for the
operator ${\bf g}$. Here and below we use the units $c=k_B=\hbar=1$.
Making a conformal transformation from the
infinite plane to the cylinder, one easily finds: 
\begin{equation}
<tr [{\bf g}(z,\bar{z})\bbox{\sigma}^z]tr [{\bf g}(0)\bbox{\sigma}^z]>
={\pi T \over \sqrt{\sin (\pi T z) \sin(\pi T \bar{z})}}.\end{equation}
The result for real time is given by a straightforward analytic
continuation, $\tau=i t$. To obtain $\chi''(q,\omega)$, one simply
performs a Fourier transform. Integration over $q$ in the limit 
$\omega \to 0$ gives NMR $T_1$\cite{Sachdev}. These results are,
of course, well known.

In order to obtain the logarithm, one has to go further\cite{Affleck,Singh}
and do a perturbative expansion in the leading irrelevant operator in
Eq.(\ref{hamilt}), and then collect the diverging terms in a  Renormalization
Group (RG) scheme\cite{Affleck1,BA}. To the first order in the leading
irrelevant operator the correction can be easily calculated. The details of a
similar calculation of this diagram in a finite-size chain at $T=0$ can be found in
Ref. \onlinecite{BA}. The first-order correction has the following form:
\end{multicols}
\widetext
\begin{equation}
\delta<tr [{\bf g}(z,\bar{z})\bbox{\sigma}^z]tr [{\bf g}(0)\bbox{\sigma}^z]>
={\pi^2 \lambda_0 T \over \sqrt{3 \sin [\pi T z] \sin [\pi T \bar{z}]}}
\left\{\ln\left[{\sin [\pi T z] \sin [\pi T \bar{z}] \over (\pi T/T_0)^2}\right]
+\hbox{const}\right\}
\label{pert}
\end{equation} 
\begin{multicols}{2}
\narrowtext
Here $\lambda_0$ is the ``bare'' coupling constant for a theory defined with 
a cutoff at $T = T_0$.

We can now sum the leading logarithmic contributions using the
standard Callan-Symanzik RG equations for the
staggered spin correlation function $\chi_s(r,\tau,T,\lambda)$: 
\begin{equation}
\label{RG}
[- \partial/\partial \ln{T} 
+\beta(\lambda)\partial/\partial \lambda + 
2 \gamma(\lambda)] \chi_s(r,r T,\tau T,  \lambda) = 0,
\end{equation}
where $\beta(\lambda)$ is the beta function for the 
coupling constant $\lambda$ in Eq.(\ref{hamilt}):
\begin{equation}
{d \lambda \over d \ln{T}} = - \beta(\lambda) 
\label{beta}
\end{equation}
and $\gamma(\lambda)$ is the anomalous dimension.  In Eq. (\ref{RG})
the $T$-derivative acts only on the first argument of $\chi_s$; $rT$ and $\tau T$ are 
held fixed. The solution of Eq.(\ref{RG}) can be written as follows:
\end{multicols} \widetext
\begin{equation}
\label{sol}
\chi_s(r, \tau, T, \lambda_0) = exp\left(- 2 \int_{\lambda_0}^{\lambda(T)}  
{\gamma[\lambda'] \over \beta(\lambda')} d \lambda' \right)
F[\lambda(T), r T, \tau T],
\end{equation}
\begin{multicols}{2} \narrowtext
where $\lambda_0 \equiv \lambda(T_0)$ is the "bare" 
coupling - a coupling at the
energy cutoff scale $T_0$, $F[\lambda(T),r T, \tau T]$ is 
an arbitrary function of the
effective coupling constant at scale $T$, $\lambda(T)$.

Since the coupling constant flows to zero as $T$ is decreased, 
one can use perturbative expressions for $\gamma(\lambda)$ and 
$\beta(\lambda)$ to determine the long-distance asymptotics for the
staggered spin susceptibility.
The universal terms in the perturbative expansion for the $\beta$-
function\cite{Solyom} and the anomalous dimension\cite{Affleck,Singh} are 
known,
\begin{eqnarray}
\label{bta}
\beta(\lambda) &=& - (4 \pi/\sqrt{3}) \lambda^2 - 
(1/2)(4 \pi/\sqrt{3})^2 
\lambda^3 \\
\label{gmma}
\gamma(\lambda) &=& 1/2 - (\pi/\sqrt{3}) \lambda.
\end{eqnarray}
Thus the effective coupling is given by:
\begin{equation} {1\over \lambda (T)}
={4\pi \over \sqrt{3}}\left\{\ln (\Lambda/T)+{1 \over 2}\ln [\ln 
(\Lambda/T)]\right\}+ O(1),\label{lambda(T)}
\end{equation}
where 
\begin{equation}
\Lambda = const \cdot \sqrt{\lambda_0} e^{\sqrt{3}/(4 \pi \lambda_0)} T_0.
\end{equation}
Thus, we can rewrite the integral in Eq.(\ref{sol}):
\begin{equation}
\label{factor}
\int_{\lambda_0}^{\lambda(T)} 
{\gamma(\lambda) \over \beta(\lambda)} d\lambda = {1 \over 2} \ln{T_0 \over T} + 
{1 \over 4} \ln{\lambda(T) \over \lambda_0} + \dots
\end{equation}
In general, one can expand the staggered spin susceptibility Eq.(\ref{sol})
in powers of $\lambda(T)$:
\end{multicols} \widetext
\begin{equation}
\chi_s(r,\tau,T,\lambda )={1\over r}\sqrt{\lambda_0\over \lambda 
(T)}e^{\sum_{n=1}^\infty a_n[\lambda (T)^n-
\lambda_0^n]}\sum_{m=0}^\infty F_m(r T, \tau T)\lambda (T)^m 
\label{GF}\end{equation}
\begin{multicols}{2} \narrowtext
The coefficients, $a_n$ and the functions $F_m(r T, \tau T)$ can 
be determined from the perturbative expansion in the leading irrelevant
operator. 

We can now RG-improve the perturbative result  Eq.(\ref{gcorr}), Eq.(\ref{pert}). 
This can be done by expanding Eq.(\ref{GF}) to first order in the bare coupling constant, 
$\lambda_0$, and comparing it with the perturbative calculations. We find:
\end{multicols} \widetext
\begin{equation}
\chi_s(r, \tau, T, \lambda) = {1 \over (2 \pi)^{3/2}} 
{\pi T \sqrt{\ln{\Lambda \over T} + {1 \over 2} \ln \left(\ln{\Lambda \over T}\right)} \over 
\sqrt{\sin(\pi T z) \sin(\pi T \bar{z})}}\left(1 + {1 \over 4 \ln(\Lambda/T)}
\ln[\sin(\pi T z) \sin(\pi T \bar{z})]\right).   
\label{reslt}
\end{equation}
\begin{multicols}{2} \narrowtext 
This expression is our final perturbative result, which has 
to be Fourier-transformed and continued analytically to real frequencies.
For this purpose it is more convenient to work with temperature-dependent
anomalous dimension $\eta(T)$ instead of the form Eq.(\ref{reslt}) with 
the logarithms. To the same order we can write:
\end{multicols} \widetext
\begin{equation}
\chi_s(r, \tau, T, \lambda) = {\pi T \over (2 \pi)^{3/2}} \sqrt{\ln{\Lambda \over T} + 
{1 \over 2} \ln \left(\ln{\Lambda \over T}\right)} 
(\sin(\pi T z) \sin(\pi T \bar{z}))^{-\eta(T) \over 2},
\label{rslt}
\end{equation}
\begin{multicols}{2} \narrowtext
where
\begin{equation}
\eta(T) = 1 - {1 \over 2 \ln{\Lambda \over T}}
\end{equation}
The analytic continuation of Eq.(\ref{rslt}) is then analogous to 
that in case of a Luttinger liquid, which is well 
known\cite{Tsvelik,Schulz,Shankar,SSS}. We therefore only discuss
quantities which can be measured by inelastic neutron scattering and
NMR. Continuing Eq.(\ref{rslt}) to real frequencies, we get:
\end{multicols} \widetext
\begin{eqnarray}
Im \chi(q,\omega) & =  & {2^{\eta(T) - 2} \over (2 \pi)^{3/2} \pi T}
\sin(\pi \eta(T)/2) 
\sqrt{\ln{\Lambda \over T} + {1 \over 2} \ln \left(\ln{\Lambda \over T}\right)}
\times \\ \nonumber
 & & \times Im\left\{B\left({i(\omega - q) \over 4 \pi T} 
+ {\eta(T) \over 4}, 1 - {\eta(T) \over 2} \right) 
B\left({i(\omega + q) \over 4 \pi T} + {\eta(T) \over 4}, 1 
- {\eta(T) \over 2} \right)\right\},
\label{imchi}
\end{eqnarray}
\begin{multicols}{2} \narrowtext
where $B(x,y) \equiv \Gamma(x) \Gamma(y)/\Gamma(x+y)$ is the beta-function.
An immediate consequence of the temperature-dependent anomalous dimension is
that the correlation length acquires additional logarithmic temperature
dependence, which can be observed in the inelastic neutron scattering 
experiments:
\begin{equation}
\xi^{-1} = \pi T\left(1 - {1 \over 2 \ln{T_0 \over T}}\right).
\end{equation}
This also agrees with thermal Bethe ansatz calculations\cite{nomura}.

Let us now compute the nuclear relaxation rates. Nuclear spins 
are coupled to electron degrees of freedom by the magnetic
hyperfine hamiltonian:
\begin{equation}
H_{HF} = \sum_{\alpha,i,j} A^{ij}_{\alpha}I_{i \alpha} S_{j \alpha},
\end{equation}
$I$ is the nuclear spin, $S$ is the electron spin, $\alpha$ enumerates
spin projections for sites $i$ and $j$ . We will use the
following expressions\cite{slichter} for calculating $T_1$ and $T_{2G}$:
\end{multicols} \widetext
\begin{eqnarray}
{1 \over T_1} &=& {2 k_B T \over  \hbar^2} \int {dq \over 2 \pi}
A_{\perp}^2(q){Im\chi(q,\omega_0) \over \omega_0} \\
\left({1 \over T_{2G}}\right)^2 &=& {p \over 8 \hbar^2} \left[
\int {dq \over 2 \pi} 
A_{\parallel}^4(q) \chi^2(q) - \left\{\int {dq \over 2 \pi}
A_{\parallel}^2(q) \chi(q) \right\}^2 \right].
\end{eqnarray}
\begin{multicols}{2} \narrowtext
Here $A_{\parallel}(q)$ and $A_{\perp}(q)$ are the hyperfine 
couplings parallel and perpendicular to the easy axis of the
crystal, $\omega_0$ is the nuclear resonance frequency, which is much
smaller than any other electron  energy scale. The magnetic 
field is directed along the $c$-axis.
The q-dependence is smooth and arises from appropriate form 
factors. The susceptibility $\chi$ should, in principle, include contributions
from both the uniform and staggered spin fluctuations. However, simple
power counting\cite{Sachdev} shows that the staggered component is dominant
at small $T$. For the purpose of comparison of our theory with experiment 
it is convenient to define normalized dimensionless
NMR rates\cite{takig}, which should be universal functions of $T/J$:
\begin{eqnarray}
(1/T_1)_{norm} &=& {\hbar J \over T_1  A_{\perp}^2(\pi)} \\
(\sqrt{T}/T_{2G})_{norm} &=& \left(k_B T \over p J\right)^{1/2}
{\hbar J \over A_{\parallel}^2(\pi) T_{2G}} 
\end{eqnarray}
A complete calculation of the NMR relaxation rates gives:
\end{multicols} \widetext
\begin{eqnarray} 
(1/T_1)_{norm} & = & 2 D \sqrt{\ln{\Lambda \over T} + {1 \over 2}
\ln\left(\ln{\Lambda \over T}\right)}\left(1 + {\ln{2} \over 2 \ln{\Lambda 
\over T}} + O\left[{1 \over \ln^2{\Lambda \over T}}\right]\right) \\
(\sqrt{T}/T_{2G})_{norm} & = &  {\sqrt{I_0} D \over 4 \sqrt{\pi}}
\sqrt{\ln{\Lambda \over T} + {1 \over 2} \ln\left(\ln{\Lambda \over T}\right)}
\left(1 + {\gamma + 4 \ln{2} + {I_1 \over 2 I_0} \over 2 \ln{\Lambda \over T}} 
+ O\left[{1 \over \ln^2{\Lambda \over T}}\right]\right).
\end{eqnarray}
\begin{multicols}{2} \narrowtext
Here $D = 1/(2 \pi)^{3/2}$ is the non-universal amplitude, 
$\gamma \simeq 0.5772157$ is the Euler's constant, while the integrals
$I_0$ and $I_1$ are given by
\begin{eqnarray}
I_0 & = & \int_0^{\infty} dx \left| {\Gamma\left({1 + ix \over 4}\right) \over
\Gamma\left({3+ix \over 4}\right)}\right|^4 \simeq 71.2766 \nonumber \\
I_1 &=&  \int_0^{\infty} dx \left| {\Gamma\left({1 + ix \over 4}\right) \over
\Gamma\left({3+ix \over 4}\right)}\right|^4 \times \nonumber \\
 & \times &Re\left[\Psi\left({1 + ix \over 4}\right) + \Psi\left({3 + ix \over 4}\right)\right] \simeq - 259.94,
\label{rates}
\end{eqnarray}
where $\Psi(x)$ is digamma function. The $1/\ln(\Lambda/T)$ term could be 
incorporated to redefine the cutoff $\Lambda$ as in Ref.\cite{Affleck1}.
Thus, up to terms $O(1/\ln^2(\Lambda/T))$ the temperature dependence 
for $1/T_1$ or $\sqrt{T}/T_{2G}$ is actually given by the square root
of the $log$ and $loglog$ terms in the numerator of Eq.(\ref{rates}). The
ratio of the relaxation rates, however, is only weakly temperature dependent.
We find:
\begin{equation}
\left(T_{2G} \over T_1 \sqrt{T}\right)_{norm} \simeq 1.680 
\left(1 + {0.7632 \over \ln(\Lambda/T)} \right) 
\end{equation}
\begin{figure}
\epsfxsize=3 in
\epsfbox{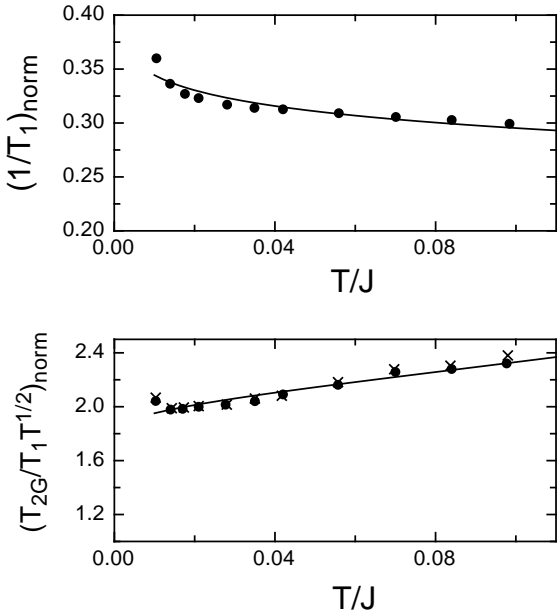} 
\caption{NMR $T_1$ and $T_{2G}/T_{1}T^{1/2}$ vs $T/J$ from Takigawa 
{\em et al.} \protect{\cite{takig}} fitted to our expression.}
\end{figure}
To summarize, the new effect of the higher-order corrections in the leading 
irrelevant operator to the dynamic spin susceptibility is, apart from the
$log log$ term in the common factor 
$\sqrt{\ln(\Lambda/T) + 0.5 \ln[\ln(\Lambda/T)]}$, the 
temperature-dependent anomalous dimension  and logarithmic corrections to
the correlation length.  These only lead to an additional
weak $O(1/\ln(\Lambda/T))$ temperature dependence for
the relaxation rates $1/T_1$ and $1/T_{2G}$, which
can be incorporated as a correction to the non-universal cutoff 
scale $\Lambda$. 
The relaxation rate ratio $(T_{2G}/T_1\sqrt{T})$, however, filters out
the common factor, and therefore picks up weak 
$1/\ln(\Lambda/T)$ temperature dependence, which we have calculated.
We note that our result is similar to the anzats used by
Starykh {\em et al}\cite{starykh}. There are, however, important
differences. Starykh {\em et al.}\cite{starykh}
don't have the $loglog$ term, which turns out to be the
most important correction in this approximation. 
The $1/\ln(T_0/T)$ weak temperature
dependence for the ratio of the relaxation rates was also not 
explicitly obtained in Ref.\cite{starykh}.
Our theoretical results are in excellent agreement with 
experimental data of Takigawa {\em et al.}\cite{takigawa,takig}
on Sr$_2$CuO$_3$, as shown in Fig.1. 

I would like to thank I. Affleck and L.P. Gor'kov for
useful discussions and comments.
This work was supported by the National High Magnetic
Field Laboratory 
through NSF cooperative agreement No. DMR-9527035 and the
State of Florida.

\end{multicols} 

\begin{references}
\bibitem{Affleck:review} For a review see I. Affleck, {\it Fields,
Strings and Critical Phenomena}
[ed. E. Br\'ezin and J. Zinn-Justin, North-Holland,
Amsterdam, 1989]; 511.
\bibitem{Affleck} I. Affleck, D. Gepner, H.J. Schulz and
T. Ziman, J. Phys.  {\bf A22},
511 (1989).
\bibitem{Singh} R.R. Singh, M.E. Fisher and R. Shankar,
Phys. Rev. {\bf B39}, 2562 (1989).
\bibitem{Affleck1}  I. Affleck, J. Phys. {\bf A31}, 4573 (1998).
\bibitem{Lukyanov} S. Lukyanov and A. Zamalodchikov,
Nucl. Phys. {\bf B493}, 571 (1997);
S. Lukyanov, Nucl. Phys. {\bf B522}, 533 (1998).
\bibitem{Kubo} K. Kubo, T.A. Kaplan and J. Borysowicz, 
Phys. Rev. {\bf B38},
11550 (1988).
\bibitem{Liang} S. Liang, Phys. Rev. Lett. {\bf 64}, 
1597 (1990).
\bibitem{Sandvik} A.W. Sandvik and D.J. Scalapino, Phys. 
Rev. {\bf B47}, 12333 (1993).
\bibitem{Hallberg} K. Hallberg, P. Horsch and G. 
Martinez, Phys. Rev. {\bf
B52}, R719 (1995).
\bibitem{Koma}T. Koma and N. Mizukoshi, J. Stat. Phys. 
{\bf 83}, 661 (1996).
\bibitem{neutrons} D. A. Tennant, R. A. Cowley, S. Nagler
and A. M. Tsvelik, Phys. Rev. B {\bf 52}, 13368 (1995).
\bibitem{takigawa} M. Takigawa, N. Motoyama, H. Eisaki and
S. Uchida, Phys. Rev. Lett. {\bf 76}, 4612 (1996)
\bibitem{starykh} O. A. Starykh, R. R. P. Singh, and A. W. Sandvik,
Phys. Rev. Lett. {\bf 78}, 539 (1997).
\bibitem{BA} V. Barzykin, I. Affleck, J. Phys. {\bf A}, {\bf 32}, 867 (1999).
\bibitem{Tsvelik} A. M. Tsvelik, {\em Quantum Field Theory
in Condensed Matter Physics} (Cambridge University Press,
Cambridge, 1995)
\bibitem{Sachdev} S. Sachdev, Phys. Rev. B {\bf 50}, 13006 (1994).
\bibitem{Solyom} J. Solyom, Adv. Phys. {\bf 28}, 201 (1979).
\bibitem{Schulz} H.J.Schulz, Phys. Rev. B {\bf 34}, 6372 (1986).
\bibitem{Shankar} R. Shankar, Int. J. Mod. Phys. B {\bf 4}, 2371 (1990).
\bibitem{SSS} S. Sachdev, T. Senthil, and R. Shankar,
Phys. Rev. B {\bf 50}, 258 (1994).
\bibitem{nomura} K. Nomura and M. Yamada, Phys. Rev. B {\bf 43}, 8217 (1991).
\bibitem{slichter}
C,\ P.\ Slichter, in {\em Strongly Correlated Electronic
Systems}, ed.\ by K.\ S.\ Bedell {\em et al}, (Addison Wesley, 1994).
\bibitem{takig} M. Takigawa, O. A. Starykh, A. W. Sandvik, and
R. R. P. Singh, Phys. Rev. B {\bf 56}, 13681 (1997).
\end{references}
\end{document}